\documentclass[]{article}

\usepackage{color}
\usepackage{graphicx}
\usepackage{mathptmx}

\begin{document}
\noindent	
{\bf \Large Effects of drug resistance in the tumour-immune system with chemotherapy
	treatment}\\
	
\noindent Jos\'e Trobia\textsuperscript{1,2,*}, Enrique C
Gabrick\textsuperscript{1}, Evandro G Seifert\textsuperscript{3}, Fernando S
Borges\textsuperscript{4}, Paulo R Protachevicz\textsuperscript{5}, Jos\'e D
Szezech Jr\textsuperscript{1,2}, Kelly C Iarosz\textsuperscript{5,6,7}, Moises S
Santos\textsuperscript{5}, Iber\^e L Caldas\textsuperscript{5}, Kun
Tian\textsuperscript{8}, Hai-Peng Ren\textsuperscript{8,9}, Celso
Grebogi\textsuperscript{8,10}, Antonio M Batista\textsuperscript{1,2,5}\\
	
{\footnotesize \noindent 
\textsuperscript{1}Graduate Program in Science - Physics, State	University of Ponta Grossa, 84030-900, Ponta Grossa, PR, Brazil. 
\textsuperscript{2}Department of Mathematics and Statistics, State University of Ponta Grossa, 84030-900, Ponta Grossa, PR, Brazil. 
\textsuperscript{3}Department of Physics, State University of Ponta Grossa, 84030-900, Ponta Grossa, PR, Brazil. 
\textsuperscript{4}Center for Mathematics, Computation, and	Cognition, Federal University of ABC, 09606-045, S\~ao Bernardo do Campo, SP,
		Brazil. \textsuperscript{5}Institute of Physics, University of S\~ao Paulo,	05508-900, S\~ao Paulo, SP, Brazil.
\textsuperscript{6}Faculty of Tel\^emaco Borba, FATEB, 84266-010, Tel\^emaco Borba, PR, Brazil. 
\textsuperscript{7}Graduate Program in Chemical Engineering Federal	Technological University of Paran\'a, Ponta Grossa, 84016-210, Paran\'a,
		Brazil. \textsuperscript{8}Shaanxi Key Lab of Complex System Control and Intelligent Information Processing, Xi'an University of Technology, Xi'an
		710048, PR China. \textsuperscript{9}Xi'an Technological University, Xi'an, 710021,	PR China. \textsuperscript{10}Institute for Complex Systems and Mathematical
	Biology, University of Aberdeen, AB24 3UE, Aberdeen, Scotland, United Kingdom.\\
}
		
	\footnotesize	Corresponding author: jtrobia@gmail.com
		
		\begin{abstract}  \noindent
Cancer is a term used to refer to a large set of diseases. The cancerous cells
grow and divide and, as a result, they form tumours that grow in size. The
immune system recognise the cancerous cells and attack them, though, it can be
weakened by the cancer. One type of cancer treatment is chemotherapy, which uses
drugs to kill cancer cells. Clinical, experimental, and theoretical
research has been developed to understand the dynamics of cancerous cells
with chemotherapy treatment, as well as the interaction between tumour growth
and immune system. We study a mathematical model that describes the cancer
growth, immune system response, and chemotherapeutic agents. The immune system
is composed of resting cells that are converted to hunting cells to combat the
cancer. In this work, we consider drug sensitive and resistant cancer cells. We
show that the tumour growth can be controlled not only by means of different
chemotherapy protocols, but also by the immune system that attacks both
sensitive and resistant cancer cells. Furthermore, for all considered
protocols, we demonstrate that the time delay from resting to hunting cells
plays a crucial role in the combat against cancer cells.

		\end{abstract}

	\normalsize
\section{Introduction}

An abnormal growth of cells can cause a malignant or cancerous tumour to invade
nearby tissues and possibly to spread to other organs \cite{hockel19}. Cancer
is a group of diseases, being a public health problem in all countries of the
world \cite{siegel15}. Many types of treatment have been developed to eliminate
cancer cells, such as surgery \cite{brennan05}, chemotherapy \cite{devita08},
and radiation \cite{ronckers04}. One of the chemotherapeutic treatments is the
immunotherapy \cite{frankel13}.

Mathematical models have been used to stu\-dy different types of cancer and
stages of tumour progression \cite{iarosz15,weerasinghe19}. In 1972, Greenspan
\cite{greenspan72} constructed a ma\-thematical model of tumour growth to
analyse the evolution of carcinoma. A model of tumour induced capillary growth
was proposed by Balding and McElwain \cite{balding85} in 1985. In the 1990s,
Tracqui et al. \cite{tracqui95} and Panetta \cite{panetta96} added chemotherapy
to study the effects of chemotherapeutic agents on spatio-temporal growth and
tumour recurrence, respectively. Recently, L\'opez et al.
\cite{lopez19a,lopez19b} formulated a model of tumour growth with cytotoxic
chemotherapeutic agents to analyse the role of dose-dense protocols.

The immune system has as its main function to protect the body against
infection and illness. It can recognise the cancerous cells and eliminate them,
though, the cancer can weaken the immunity \cite{gonzalez18}. The cancer
treatment that takes advantage of the immune system is known as immunotherapy
\cite{farkona16}. Some therapies based on the immune system consists of
monoclonal antibodies, vaccines, and T-cell transfer \cite{oiseth17}.
Mathematical and computational studies of cancer immunotherapy have been
performed to understand the interactions between immunity and tumour growth
\cite{nani00,konstorum17}. Borges et al. \cite{borges14} presented a
tumour-immune model with chemotherapy treatment. They considered a time-delay
between the conversion from resting to hunting cells, the main immune system
reaction. Ren et al. \cite{ren17} demonstrated analytical result for impulse
che\-motherapy parameter to eliminate the cancer cells.

Cancers can develop resistance to chemotherapeutic agents \cite{housman14}. Drug
resistance is a phenomenon that occurs when cancer cells are unaffected by
chemotherapy. Experiments have yielded information about the mechanisms of
cancer drug resistance \cite{gottesman02}. Sun et al. \cite{sun16} modelled
drug sensitive and resistant cancer cells in response to chemotherapeutic
treatment. Trobia et al. \cite{trobia20} created a model of brain tumour growth
with drug resistance. They demonstrated that the time interval of the drug
application plays an important role in the treatment to eliminate the cancerous
cells.

In this work, we include drug resistance in the tu\-mour-immune model proposed
by Borges et al. \cite{borges14} and analyse its effect on the system. In our
mathematical model, the immune system is composed of resting and hunting cells,
while the cancer is separated into drug sensitive and drug resistant cells. We
consider chemo\-therapy to combat the tumour growth. However, the
che\-motherapeutic agents also attack the immune system. We show that the
tumour growth can be controlled by means of different chemotherapy protocols.

The paper is organised as follows. In Section $2$, we introduce the
tumour-immune system with drug resistance. Section $3$ presents our results
about the effects of the drug resistance. In the last Section, we draw
our conclusions.


\section{Mathematical model}

Cancer drug resistance has been a difficulty in chemo\-therapy cancer treatment
\cite{vasan19}, and the challenge is how to identify and avoid the resistance
\cite{godefridus18}. Many resear\-chers have carried out tests to find new
strategies in the treatment of tumours associated with drug resistance
\cite{pascual19}. 

\begin{figure}[hbt!]
	\centering
	\includegraphics[scale=0.2]{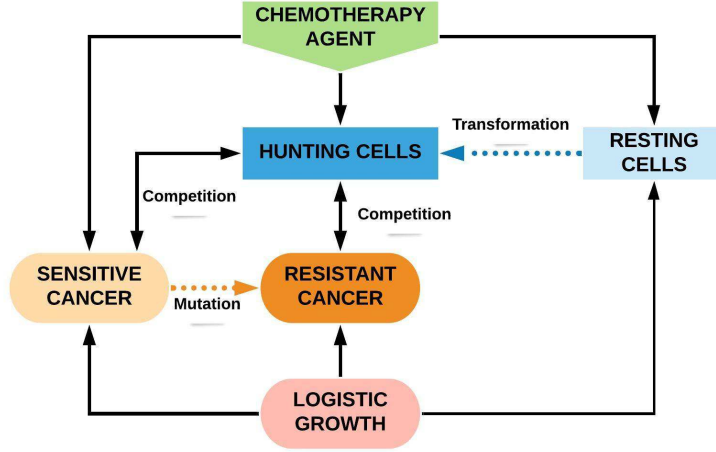}
	\caption{(Colour online) Schematic representation of the model.} 
	\label{fig1}
\end{figure}

In this work, we proposed a mathematical model that describes cancerous cell
growth, where we include cancer drug resistance. The cancer cells are separated
into sensitive and resistant cells, as they are attacked by the immune system.
In the immune system, the resting cells are converted to hunting cells. We
consider that the cancerous and resting cells have a logistic growth, while the
hunting cells have a form of programmed cell death, known as apoptosis. The
chemotherapeutic agent is applied to kill the cancer, it affects all cells,
except the drug resistant cancer cells, as shown in Fig. \ref{fig1}.

 The model is given by
\begin{eqnarray}
\frac{dC_S}{dt} & = & q_1C_S\left(1-\frac{C_S+C_R}{K_1}\right)-\alpha_1C_SH -uF[Z]C_S-\frac{p_1C_SZ}{a_1+C_S}, \\
\frac{dC_R}{dt} & = & q_2C_R\left[1-\frac{C_S+C_R}{K_1}\right]-\alpha_2C_RH +uF[Z]C_S, \\
\frac{dH}{dt} & = & \beta_1HR\left(t-\tau\right)-d_1H-\alpha_3H
\left[C_S+C_R\right] -\frac{p_2HZ}{a_2+H}, \\
\frac{dR}{dt} & = & q_3R\left(1-\frac{R}{K_2}\right)-\beta_1HR(t-\tau) -\frac{p_3RZ}{a_3+R},  \\
\frac{dZ}{dt} & = & \Phi-\left(\zeta+\frac{g_1C_S}{a_1+C_S}+\frac{g_2H}{a_2+H}+\frac{g_3 R}{a_3+R}\right)Z, 
\end{eqnarray}
where $C_{S}$ and $C_{R}$ are the concentration of drug sensitive and resistant
cells (kg.m$^{-3}$), respectively, $H$ is the concentration of hunting cells
(kg.m$^{-3}$), $R$ is the concentration of resting cells (kg.m$^{-3}$), $Z$ is
the concentration of the chemotherapeutic agent (mg.m$^{-2}$), $t$ is the time
(day), $\tau$ is the delay time from resting to hunting cells, and $F(Z)$ is a
function defined as
\begin{equation}
F(Z)=\left\{
\begin{array}{rcl}
0,  &\mbox{} & Z=0 \\  
1,  &\mbox{} & Z>0
\end{array}
\right. . 
\end{equation}
Besides that, $p_i$ represents the predation coefficient of the
chemothe\-rapeutic agent, $a_i$ corresponds to the rate at which the cells
achieve the carrying capacity when there is no competition and predation, and
$g_i$ represents the combination rates of the chemotherapeutic agent with the
cells \cite{pinho02}. The parameters $p_i$ and $g_i$ are related with the
strength of the Holling type 2 interaction functions. Holling \cite{holling65}
proposed types of functional responses to different types of interactions. The
type 2 function describes the response of many interacting predators and has
the characteristics of decelerating the intake rate. The parameter values that
we use in our simulations are given in Table \ref{tab1}.

\begin{table}[htbp]	\centering
	{
		\centering
		\caption{Parameters values according to the literature.}
		\label{tab1}	\centering
		\begin{tabular}{c|c|c} 
			\hline
			Parameter & Values & Description \\ \hline
			$q_1$ & $0.18$ day$^{-1}$ & Proliferation \\
			$q_2$ & $0.18$ day$^{-1}$ & rate \cite{siu86,banerjee08} \\
			$q_3$ & $0.0245$ day$^{-1}$ &  \\ \hline
			$d_1$ & $0.0412$ day$^{-1}$ & Death rate \cite{kuznetsov94}  \\ \hline
			$\beta_1$ & $6.2\times 10^{-9}$ (cells $\cdot$ day)$^{-1}$ & Conversion rate
			\cite{kuznetsov94} \\ \hline
			$\Phi$ & $0-200$ mg(m$^{2}$.day)$^{-1}$ & Chemotherapy
			\cite{stupp05,strik12}   \\ \hline
			$\zeta$ & $0.2$ day$^{-1}$ & Absorption rate \cite{borges14} \\ \hline
			$u$ & $10^{-3}$ day$^{-1}$ & Mutation rate \cite{trobia20} \\ \hline
			$\alpha_{1},\alpha_{2}$ & $1.101\times 10^{-7}$ (cells $\cdot$ day)$^{-1}$ &
			Competition \\
			$\alpha_{3}$ & $3.422\times 10^{-10}$ (cells $\cdot$ day)$^{-1}$ & coefficients
			\cite{kuznetsov94} \\ \hline
			$K_1$ & $5\times 10^{6}$ cells & Carrying  \\
			$K_2$ & $1\times10^{7}$ cells & capacity \cite{siu86,banerjee08} \\ \hline
			$\tau$ & $45.6$ days & Time delay \cite{banerjee08} \\ \hline
		\end{tabular}
	}
\end{table}

We introduce the following dimensionless variables $c_s=C_S/K_T$, $c_r=C_R/K_T$,
$h=H/K_T$, $r=R/K_T$, and $z=\zeta Z$, where $K_T=K_1+K_2$ and $t^*=t/day$. We
consider $K_1^*=K_1/K_T$, $K_2^*=K_2/K_T$, $u^*=u$ day, $d_1^*=d_1$ day,
$\beta^*_1=\beta_1 K_T$ day, $\Phi^*=\Phi$ day, $\zeta^*=\zeta$ day, $q^*_i=q_i$
day, $\alpha^*_i=\alpha_iK_T$ day, $p_i^*=p_i/(\zeta K_T)$ day, $g^*_i=g_i$ day,
and $a_i^*=a_i/K_T$ ($i=1,2,3$). Redefining the variables and removing the
stars, we obtain

\begin{eqnarray}
\frac{dc_s}{dt} & = & q_1c_s\left(1-\frac{c_s+c_r}{K_1}\right)-\alpha_1c_sh-uF[z]c_s -\frac{p_1c_sz}{a_1+c_s},  \\
\frac{dc_r}{dt} & = & q_2c_r\left[1-\frac{c_s+c_r}{K_1}\right]-\alpha_2c_rh +uF[z]c_s, \\
\frac{dh}{dt} & = & \beta_1hr\left(t-\tau\right)-d_1h-\alpha_3h
\left[c_s+c_r\right] -\frac{p_2hz}{a_2+h},  \\
\frac{dr}{dt} & = & q_3r\left(1-\frac{r}{K_2}\right)-\beta_1hr(t-\tau) -\frac{p_3rz}{a_3+r},\\
\frac{dz}{dt} & = & \Phi\zeta-\left(\zeta+\frac{g_1c_s}{a_1 +c_s}+
\frac{g_2 h}{a_2+h}\right.  \left.+\frac{g_3 r}{a_3+r}\right) z, 
\end{eqnarray}
The dimensionless parameter values are given in Table \ref{tab2}.

\begin{table}[htbp]
	\begin{center}
		\caption{Dimensionless parameters.}
		\label{tab2}
		\begin{tabular}{c|c} 
			\hline
			Parameter & Values \\ \hline
			$q_{1}$ & $0.18$ \\ \hline
			$q_{2}$ & $0.18$ \\ \hline
			$q_{3}$ & $0.0245$ \\ \hline
			$d_1$ & $0.0412$ \\ \hline
			$\beta_1$ & $9.3\times 10^{-2}$ \\ \hline
			$p_1$ & $1\times10^{-3}$ \\ \hline
			$p_2$ & $1\times10^{-3}$ \\ \hline
			$p_3$ & $1\times10^{-3}$ \\ \hline
			$a_1$ & $1\times10^{-4}$ \\ \hline
			$a_2$ & $1\times10^{-4}$ \\ \hline
			$a_3$ & $1\times10^{-4}$ \\ \hline
			$g_1$, $g_2$, $g_3$ & $0.1$ \\ \hline
			$\alpha_{1},\alpha_{2}$ & $1.6515$ \\ \hline
			$\alpha_{3}$ & $5.133\times 10^{-3}$ \\ \hline
			$K_1$ & $1/3$ \\ \hline
			$K_2$ & $2/3$ \\ \hline
	\end{tabular}\end{center}	
\end{table}

\section{Tumour drug resistance}

Many different powerful chemicals and clinical protocols have been used to
eliminate a wide variety of cancers. In this work, we consider both continuous
and pulsed chemotherapy treatments. In our simulations, the initial conditions
are given by $c_s(0)=0.18$, $c_r(0)=0.0$, $h(0)=0.01$, $r(0)=0.48$, and
$z(0)=0.0$.

\subsection{Continuous chemotherapy treatment}

\begin{figure}[hbt!]
	\centering
	\includegraphics[scale=0.31]{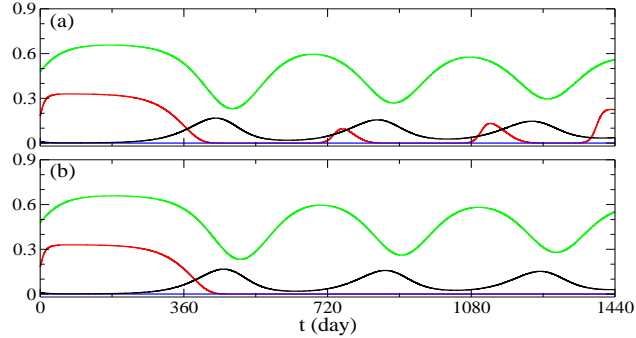}
	\caption{(Colour online) Time evolution of $c_s$ (red line), $c_r$ (blue
		line), $h$ (black line), and $r$ (green line) for (a) $\Phi=0.02$ and
		(b) $\Phi = 0.025$.} 
	\label{fig2}
\end{figure}

Figure \ref{fig2} displays the behaviour of the time evolution of $c_s$ (red
line), $c_r$ (blue line), $h$ (black line), and $r$ (green line) when
there is no cancer drug resistance ($u=0$) for a continuous chemotherapy
treatment. Increasing the value of chemotherapy dose $\Phi$ from $0.02$ (Fig.
\ref{fig2}(a)) to $0.025$ (Fig. \ref{fig2}(b)), we observe that the cancer (red
line) is killed while the cells of the immune system (black and green lines)
remain alive. 

\begin{figure}[hbt!]
	\centering
	\includegraphics[scale=0.31]{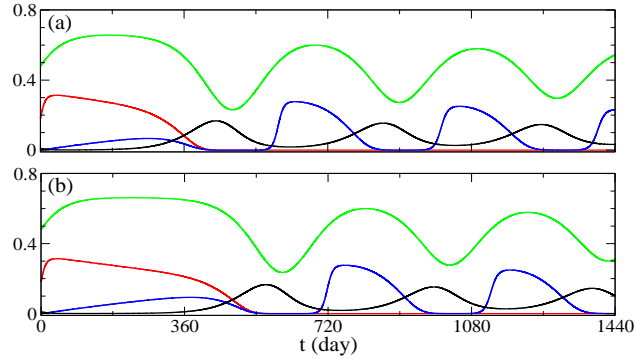}
	\caption{(Colour online) Time evolution of $c_s$ (red line), $c_r$ (blue
		line), $h$ (black line), and $r$ (green line) for $u=0.001$, (a)
		$\Phi=0.02$, and (b) $\Phi = 0.035$.} 
	\label{fig3}
\end{figure}

Drug resistance is one of the many problems in the cancer therapy. This
phenomenon is considered in our model when the mutation rate $u>0$. In Fig.
\ref{fig3}, we see the appearance of drug resistant cancer cells (blue line)
due to $u=0.001$. Increasing $\Phi$ from $0.02$ (Fig. \ref{fig3}(a)) to $0.035$
(Fig. \ref{fig3}(b)), we verify a temporary cancer remission ($c_s(t)<0.0009$
and $c_r(t)<0.0009$) for $t$ equal to $434$ and $557$ days, respectively. The
sensitive cancer cells are suppressed by the chemotherapy and the immune system.
However, the immune system by itself is not sufficient to suppress the
resistant cancer cells.

\begin{figure}[hbt!]
	\centering
	\includegraphics[scale=0.6]{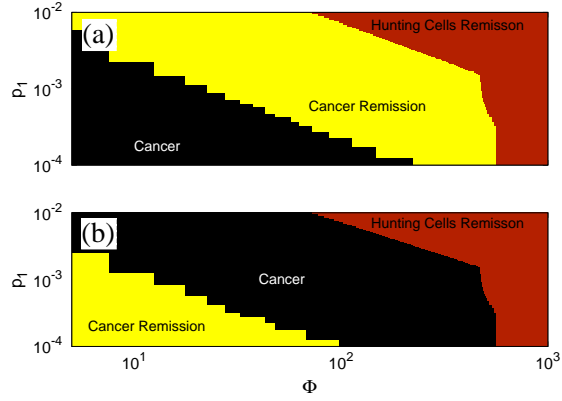}
	\caption{(Colour online) Parameter space $p_1\times\Phi$ for (a) $u=0.0$ and
		(b) $u=0.001$. We separate into three regions: cancer growth ($c_s>0.001$,
		black), cancer remission ($c_s<0.001$ and $0.0\leq c_r<0.01$, yellow),
		and hunting cells remission ($h<0.001$, red).} 
	\label{fig4}
\end{figure}

We compute the parameter space $p_1\times\Phi$ to identify the regions in which
the cancer remission occurs. Figure \ref{fig4}(a) displays the situation without
drug resistance, namely for $u=0$. We separate into three regions: cancer
grow\-th ($c_s>0$), cancer remission ($c_s<0.001$), and hunting cells remission
($h<0.001$). The cancer grows for small values of $p_1$ and $\Phi$ (black
region), but it is suppressed for larger values (yellow region). Higher values
of these parameters not only lead to the killing cancerous cells, but also
weaken the immune system with the remission of the hunting cells (red region).
When there is drug resistance, the temporary cancer remission for $100$ days
($0.0\leq c_r<0.01$) is observed for small values of $p_1$ and $\Phi$ for
$u=0.001$, as shown in Fig. \ref{fig4}(b) (yellow region). 

\subsection{Pulsed chemotherapy treatment}

Pulsed administration of chemotherapeutic drugs, also known as intermittent
therapy, is a clinical protocol in which the drug is administered and followed
by a rest period. In our simulations, we use periodically pulsed chemotherapy
and analyse different rest periods to find cancer remission.

\begin{figure}[hbt!]
	\centering
	\includegraphics[scale=0.3]{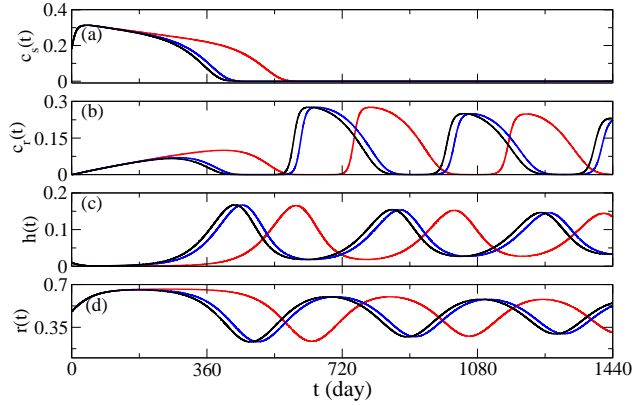}
	\caption{(Colour online) Time evolution of (a) $c_s$, (b) $c_r$, (c) $h$, and
		(d) $r$ for $\Phi=0.2$ and $u=0.001$. We consider the protocols (days of
		administration$\times$ time interval): $5\times 23$ (red line), $2\times 15$
		(blue line), and $1\times 10$ (black line).} 
	\label{fig5}
\end{figure}

Figures \ref{fig5}(a), \ref{fig5}(b), \ref{fig5}(c), and \ref{fig5}(d), exhibit
the time evolution of (a) $c_s$, (b) $c_r$, (c) $h$, and (d) $r$, respectively,
for $\Phi=0.2$, $u=0.001$, and different protocols (days of administration
$\times$ time interval). We do not observe a significant difference between the
protocols $2\times 15$ (blue line) and $1\times 10$ (black line). However, 
both are better than the protocol $5\times 23$ (red line), due to the fact that
the times for suppression and remission of $c_s$ and $c_r$, respectively, are
shorter than $5\times 23$. The suppression of $c_s$ occurs for $t$ approximately
equal to $625$ for $5\times 23$, and about $500$ for $2\times 15$ and
$1\times 10$. The temporary remission ($c_r<0.1$) starts approximately $615$
days after the chemotherapy treatment according to the protocol $5\times 23$,
and about $450$ days for the protocols $2\times 15$ and $1\times 10$.

\begin{figure}[hbt!]
	\centering
	\includegraphics[scale=0.65]{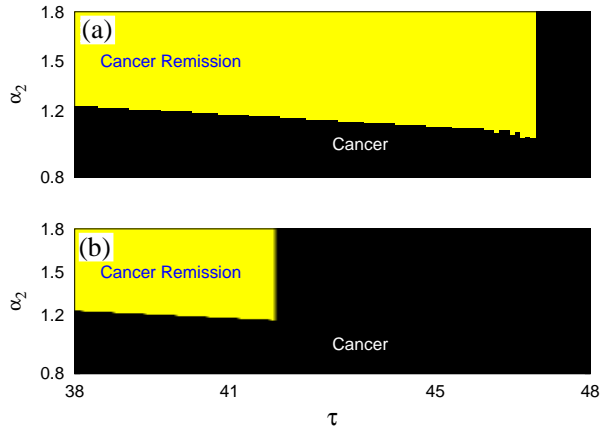}
	\caption{(Colour online) Parameter space $\alpha_2\times\tau$ for the protocol
		$5\times 23$, (a) $\Phi=0.2$, and (b) $\Phi=0.25$. The yellow and black regions
		correspond to the cancer remission and cancer.} 
	\label{fig6}
\end{figure}

The immune system plays an important role in the combat against the cancer.
Thereby, to analyse the influence of the hunting cells on the resistant cancer,
we vary the competition coefficient between the hun\-ting and drug resistant
cancer cells ($\alpha_2$), and the delay time from resting to hunting cells
($\tau$). Figure \ref{fig6} exhibits the parameter space $\alpha_2\times\tau$
for the protocol $5\times 23$, where we consider cancer remission when
$c_s<0.0001$ and $c_r<0.1$ (yellow region) for at least $100$ days, and cancer
for $c_s\geq 0.0001$ and $c_r\geq 0.1$ (black region). We verify that increasing
$\Phi$ from $0.2$ (Fig. \ref{fig6}(a)) to $0.25$ (Fig. \ref{fig6}(b)) the
cancer remission region decreases. Therefore, for larger $\Phi$ value, the
cancer remission is obtained for smaller $\tau$ value.

\begin{figure}[hbt!]
	\centering
	\includegraphics[scale=0.65]{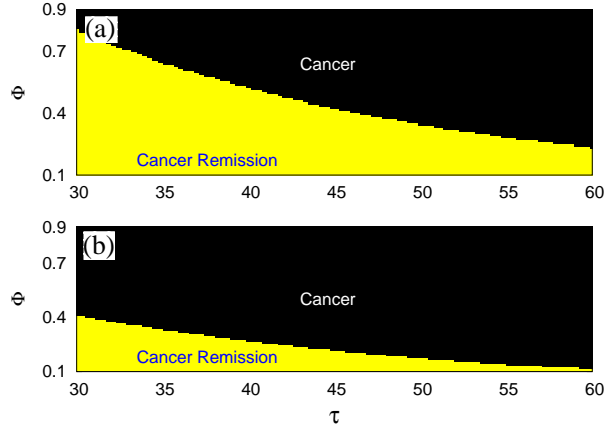}
	\caption{(Colour online) Parameter space $\Phi\times\tau$ for the protocols
		(a) $1\times 10$ and (b) $5\times 23$. The yellow and black regions correspond
		to the cancer remission and cancer.} 
	\label{fig7}
\end{figure}

We also compute the parameter space $\Phi\times \tau$ for the protocols
$1\times 10$ and $5\times 23$, as shown in Figs. \ref{fig7}(a) and
\ref{fig7}(b), respectively. Comparing Fig. \ref{fig7}(a) and Fig.
\ref{fig7}(b), we see that not only $\Phi$ and $\tau$ are important, but also
the type of protocol is relevant to increase the cancer remission region.
The cancer remission region is smaller for $5\times 23$ than $1\times 10$.

\section{Conclusions}

Drug resistance is responsible for a vast majority of cancer deaths and it is
one of the major challenges in che\-motherapy treatment. Initially some cancers
are su\-sceptible to chemotherapeutic agents, however over ti\-me they can
become resistant. Due to this fact, strategies have been used to eliminate
resistant cancer cells.

In this work, we study the effects of the drug resistance in the tumour-immune
system with chemotherapy treatment. The immune system is composed of resting
cells that can transform into hunting cells. We separate the cancer into drug
sensitive and drug resistant cells. In our simulations, we consider continuous
and pulsed chemotherapy treatment.

In the continuous chemotherapy treatment, we ve\-rify that cancer remission is
possible for smaller values of the chemotherapy intensity and the coefficient
of chemotherapeutic agent on the sensitive cancer cells. The sensitive cells
are eliminated, while the resistant cells are responsible for the remission.
With regard to the pulsed chemotherapy, we analyse three types of protocols
(days of administration $\times$ time interval): $5\times 23$, $2\times 15$, and
$1\times 10$. The protocols $2\times 15$ and $1\times 10$ exhibit almost the
same results. In both protocols, the time for the elimination of sensitive
cancer cells and the beginning of the temporary remission are less than the
protocol $5\times 23$. Furthermore, for all protocols, we show that the time
delay from resting to hunting cells plays a crucial role in the combat against
cancer cells.

Our results are in agreement with recent experimental findings related to
chemo-immunotherapy. In 2020, Roemeling et al. \cite{roemeling20} carried out
treatments to induce immune response against a type of brain tumour. They
reported a therapeutic modulation that is able to generate potent hunting cells.
In our model, the hunting cell efficiency is increased by means of the
competition coefficient between hunting cells and cancer in which the hunting 
cells kill the cancerous cells. Maletzki et al. \cite{maletzki20} in 2019
demonstrated that the combination of immune-stimulating vaccination and
cytotoxic therapy can improve long-term survival. Depending on the protocol,
they observed tumour free in mice from $25$ to $65$ weeks. In our simulations,
the tumour free occurs about $25$ weeks. Nevertheless, for small time delay
from resting to hunting cells in our model, it is possible to use different
protocols aiming to maximise the tumour free time.


\section*{Acknowledgement}
This study was possible by partial financial support fr\-om the following
Brazilian government agencies: Fun\-da\c c\~ao Arauc\'aria, National Council
for Scientific and Technological Development, Coordination for the Improvement
of Higher Education Personnel, and S\~ao Pa\-ulo Research Foundation
(2015/07311-7, 2017/18977-1, 2018/03211-6, 2020/04624-2).


\end{document}